\documentclass[runningheads]{llncs}
\usepackage[utf8]{inputenc}

\usepackage{graphicx}
\usepackage[dvipsnames]{xcolor}
\usepackage{tikz}
\usetikzlibrary{shapes}
\usetikzlibrary{arrows.meta}
\usetikzlibrary{positioning}
\usepackage{enumerate}
\usepackage{enumitem}
\usepackage{hyperref}
\definecolor{citegreen}{rgb}{0,0.5,0.2}
\hypersetup{
    colorlinks=true,
    linkcolor=blue,
    filecolor=magenta,
    urlcolor=cyan,
    citecolor=citegreen,
}

\usepackage[
style=alphabetic,
sorting=nyt,
hyperref=true, %
giveninits=true,%
backend=biber,
natbib=true,
url=false,
isbn=false,
maxnames=999, 
maxalphanames=10, 
doi=false]{biblatex}
\usepackage{caption} 
\usepackage{subcaption}
\usepackage{bussproofs}

\usepackage{stmaryrd}
\usepackage{amsmath}
\usepackage{amsfonts}
\usepackage{stmaryrd}
\usepackage{bm} 
\usepackage{circledsteps}
\usepackage{skull}
\usepackage{cancel}
\usepackage{soul} 

\usepackage{xspace}

\newtheorem{notationinner}{Notation}
\newenvironment{notation}
{\begin{notationinner}\normalfont}
    {\end{notationinner}}
\newtheorem{factinner}{Fact}
\newenvironment{fact}
{\begin{factinner}\normalfont}
{\end{factinner}}

\usepackage{float}
\floatstyle{boxed}
\restylefloat{figure}

\usepackage{listings} 
\newcommand{\listingsttfamily}{\fontfamily{SourceCodePro-TLF}\small}

\lstdefinestyle{forest}{
    language=C,
    morekeywords={INC, DEC, fi, then , true, false, push, pop, or, and, for, to, do, od, skip, rof, until, maybe, from, loop,  loopd, loopu , assert, endif, endloop, popAux, morf , ot},  
    backgroundcolor=\color{backcolour},
    basicstyle=\footnotesize\listingsttfamily,
    breakatwhitespace=false,         
    breaklines=true,                 
    belowcaptionskip=.5\baselineskip, 
    captionpos=b,                    
    commentstyle=\color{purple!40!black},
    deletekeywords={...},            
    escapeinside={(*}{*)},          
    extendedchars=true,              
    identifierstyle=\color{blue},
    firstnumber=0,                
    stepnumber=1,                
    frame=trbl,
    keepspaces=true,                 
    keywordstyle=\color{black},
    numbers=left,                    
    numbersep=7pt,                   
    numberstyle=\tiny\color{mygray}, 
    rulecolor=\color{black},         
    showspaces=false,                
    showstringspaces=false,          
    showtabs=false,                  
    stepnumber=2,                    
    tabsize=2,	                   
    title=\lstname,                   
    xleftmargin=\parindent,
}

\newcommand{\FF}[1]{{\normalfont\lstinline[style=forest]|#1|}}
\usepackage{listings} 
\lstdefinestyle{loopexit}{
    language=C,
    morekeywords={INC, DEC, fi, then , true, false, push, pop, or, and, for, to, do, od, skip, rof, until, maybe, from, loop,  loopd, loopu , assert, endif, endloop, exit, popAux},  
    backgroundcolor=\color{backcolour},
    basicstyle=\footnotesize\listingsttfamily,
    breakatwhitespace=false,         
    breaklines=true,                 
    belowcaptionskip=.5\baselineskip, 
    captionpos=b,                    
    commentstyle=\color{purple!40!black},
    deletekeywords={...},            
    escapeinside={(*}{*)},          
    extendedchars=true,              
    identifierstyle=\color{blue},
    firstnumber=0,                
    stepnumber=1,                
    frame=trbl,
    keepspaces=true,                 
    keywordstyle=\color{black},
    numbers=left,                    
    numbersep=7pt,                   
    numberstyle=\tiny\color{mygray}, 
    rulecolor=\color{black},         
    showspaces=false,                
    showstringspaces=false,          
    showtabs=false,                  
    stepnumber=2,                    
    tabsize=2,	                   
    title=\lstname,                   
    xleftmargin=\parindent,
}

\newcommand{\LL}[1]{{\normalfont\lstinline[style=loopexit]|#1|}}

\newcommand{\lead}[1]{l(#1)}
\newcommand{\clean}[2]{c(#1,#2)}

\newcommand{\bbot}{\bm{\bot}}
\newcommand{\bhsigma}{\bm{\hat{\sigma}}}

\newcommand{\bsigma}{\bm{\sigma}}
\newcommand{\bhmu}{\bm{\hat{\mu}}}

\newcommand{\bmu}{\bm{\mu}}

\newcommand{\bhnu}{\bm{\hat{\nu}}}
\newcommand{\bnu}{\bm{\nu}}

\newcommand{\bsigmap}{\bm{\sigma'}}

\newcommand{\bhtau}{\bm{\hat{\tau}}}
\newcommand{\bhtaup}{\bm{\hat{\tau}'}}
\newcommand{\btau}{\bm{\tau}}

\newcommand{\ZZ}{\mathbb{Z}}
\newcommand{\NN}{\mathbb{N}}
\newcommand{\EXPRZ}{\mathcal{Z}}

\newcommand{\EXPRB}{\mathcal{B}}
\newcommand{\WDOM}[1]{{\normalfont\operatorname{WDom}(}#1{\normalfont)}}
\newcommand{\DOM}[1]{{\normalfont\operatorname{Dom}(}#1{\normalfont)}}

\newcommand{\MIN}[1]{{\normalfont\operatorname{min}(}#1{\normalfont)}}

\newcommand{\forEscape}{\textsf{For$_{\textsf{est}}$}}
\newcommand{\forest}{{\normalfont\forEscape}\xspace}

\newcommand{\forestP}{\mathcal{F}}
\newcommand{\forestPext}{\mathcal{F}_{\operatorname{\textit{ext}}}}

\newcommand{\JANUS}{{\normalfont\textsf{Janus}}\xspace}
\newcommand{\SRL}{{\MSRL}\xspace}

\newcommand{\PRF}{{\normalfont\textsf{PRF}}\xspace}
\newcommand{\LOOP}{{\normalfont\textsf{LOOP}}\xspace}

\newcommand{\MSRL}{{\normalfont\textsf{M-SRL}}\xspace}

\newcommand{\STOF}[2]{{\normalfont\llbracket#1\rrbracket_{#2}}}
\newcommand{\STOFI}[1]{{\normalfont\llbracket#1\rrbracket}}
\newcommand{\loopExit}{{\normalfont\textsf{Loop$_{\textsf{exit}}$}}\xspace}

\definecolor{codegreen}{rgb}{0,0.6,0} 
\definecolor{codegray}{rgb}{0.5,0.5,0.5} 
\definecolor{codepurple}{rgb}{0.58,0,0.82} 
\definecolor{backcolour}{rgb}{0.95,0.95,0.90} 
\definecolor{eclipseOrange}{RGB}{200,48,0} 
\definecolor{eclipseBlue}{RGB}{0,0,172} 
\definecolor{keywordcolor}{rgb}{0.7, 0.1, 0.1} 
\definecolor{commentcolor}{rgb}{0.4, 0.4, 0.4} 
\definecolor{symbolcolor}{rgb}{0.0, 0.1, 0.6}  
\definecolor{sortcolor}{rgb}{0.1, 0.5, 0.1}    
\definecolor{aawhite}{rgb}{0.97,0.97,0.97}
\definecolor{awhite}{rgb}{0.90,0.90,0.90}
\definecolor{lgreen}{rgb}{0.94,1.0,0.98}
\definecolor{dgreen}{rgb}{0.0,0.3,0.1}
\definecolor{sgreen}{rgb}{0.0,0.7,0.3}
\definecolor{lgreen}{rgb}{0.94,1.0,0.98}
\definecolor{bgreen}{rgb}{0.00,0.50,0.25}
\definecolor{dblue}{rgb}{0.0,0.1,0.6}
\definecolor{lorange}{rgb}{1, .85, .60}
\definecolor{lblue}{rgb}{.80, .90, .95}
\definecolor{mixed}{rgb}{0.0,0.3,0.3}
\definecolor{dred}{rgb}{0.6,0.2,0.0}
\definecolor{sred}{rgb}{0.7,0.2,0.0}
\definecolor{ddred}{rgb}{0.3,0.1,0.0}
\definecolor{turq}{rgb}{0.28,0.82,0.80}
\definecolor{lyellow}{rgb}{1.00,0.97,0.94}
\definecolor{mygreen}{rgb}{0,0.6,0}
\definecolor{mygray}{rgb}{0.5,0.5,0.5}
\definecolor{mymauve}{rgb}{0.58,0,0.82}
\definecolor{codegreen}{rgb}{0,0.6,0}
\definecolor{codegray}{rgb}{0.5,0.5,0.5}
\definecolor{codepurple}{rgb}{0.58,0,0.82}
\definecolor{backcolour}{rgb}{0.95,0.95,0.92}

\begin{document}

\title{Algorithmically expressive, always-terminating model for reversible computation}
%
\titlerunning{Expressive and terminating reversible computational model}
%
\author{Matteo Palazzo\inst{1}\orcidID{0009-0008-8455-8242} \and
Luca Roversi\inst{1}\orcidID{0000-0002-1871-6109}}
\authorrunning{M. Palazzo, L. Roversi}
%
\institute{
Università degli Studi di Torino, Dipartimento di Informatica, Italy\\
\email{\{matteo.palazzo,luca.roversi\}@unito.it}}
\maketitle              
\begin{abstract}
Concerning classical computational models able to express all the Primitive Recursive Functions (\PRF), there are interesting results regarding limits on their algorithmic expressiveness or, equivalently, efficiency, namely the ability to express algorithms with minimal computational cost. By introducing the reversible programming model \forest, at our knowledge, we provide a first study of analogous properties, adapted to the context of reversible computational models that can represent all the functions in \PRF.
Firstly, we show that \forest extends Matos' linear reversible computational model \MSRL, the very extension being a guaranteed terminating iteration that can be halted by means of logical predicates.
The consequence is that \forest is \PRF-complete, because \MSRL is.
Secondly, we show that \forest is strictly algorithmically  more expressive than \MSRL: it can encode a reversible algorithm for the minimum between two integers in optimal time, while \MSRL cannot.
\end{abstract}

\section{Introduction}
\label{section:Introduction}
In relation to classical computational models results were proven regarding their algorithmic expressiveness, or efficiency, understood as the ability to express algorithms with minimal computational cost.
\par
Colson and others~\cite{Colson:AMAI96,Colson:TCS98} studied the efficiency of Primitive Recursive Functions (\PRF), proving their \emph{ultimate obstinacy property}. It means that many algorithms cannot be efficiently implemented by any term of \PRF. Among them there are the algorithms to find the \textit{minimum} between two values, which the literature see as a \emph{least standard benchmark} to argue about the efficiency of a given computational model.
\par
Matos~\cite{MATOS:TCS15} proves an analogous of \emph{ultimate obstinacy property} for
Meyer/Ritchie's \LOOP~\cite{MeyerRitchie:ACM67}, imperative computational model that characterizes \PRF. \LOOP is \PRF-correct and complete, representing all and only elements in \PRF. Roughly, \LOOP is ``obstinate'' because its iterations cannot be interrupted as soon as necessary. They must unfold to their end, no matter the state they must produce as a result becomes available in the course of the unfolding. Matos shows how to tame \LOOP ``obstinacy'' by extending \LOOP with conditional breaks and decrements, making the formalism non structured, however.

\subsubsection*{Motivations.} If the expressiveness of a computational model is valuable in the classical setting, we think it holds even greater value if the goal is to define interesting and terminating algorithms for compression/decompression, or encryption/decryption in a reversible computational setting.

\begin{lstlisting}[style=forest, caption=Term \FF{minPos} in \forest computing the function minimum in $\NN$, label=listing:Minimum minimumF in forest]
// (*$m, n \geq 0$*), x=(*$m$*), y=(*$n$*), i=0, min=0, found=0
min += x;
from ((i=0) or 0) to ((i=x) or (found=1)) {
    if (i=y) {
        min -= x;
        min += y;
        found += 1
    } else {skip}
}
// min=(*$\MIN{m,n}$*)
\end{lstlisting}

\subsubsection*{Contributions.}
Matos' \emph{linear} \emph{reversible} computational model \MSRL \cite{Matos:TCS03} is the natural counterpart of \LOOP in a reversible setting. \MSRL is \PRF-complete \cite{MatosPR:RC20}, and \PRF-correct essentially because every instance of its iterative construct `$\FF{for}\ r\ \FF{\{}P\FF{\}}$' unfolds as many times as the initial value of $r$.
\par
Inspiring to \cite{MATOS:TCS15}, we argue about why \MSRL cannot encode at least the algorithm determining the minimum between two integer numbers.
\par
To overcome \MSRL limitation, we introduce the computational model \forest, which we show it is: (i) always-terminating; (ii) reversible; (iii) able to simulate every \MSRL program, namely every \PRF function; (iv) strictly more algorithmically expressive than \MSRL.
\par
Point (iv) here above means that we can write \emph{at least} Listing~\ref{listing:Minimum minimumF in forest} in \forest which \emph{always computes} the \textit{minimum} between two naturals $m, n$ \emph{efficiently}, namely in a number of steps of order equal to the least between $m$, and $n$. In fact, we will see that \forest can compute the minimum for \emph{every pair of integers}.
\par
All achievements are possible because the iterative construct of \forest is:
\begin{align}
\label{align:from-to forest introduction}
&\FF{from (}i\FF{=}e_u\,\FF{or}\,e_{\operatorname{\textit{in}}}\FF{) to (}i\FF{=}e_v\,\FF{or}\,e_{\operatorname{\textit{out}}}\FF{)\{} P \FF{\}}
\enspace ,
\end{align}
\noindent
where $i$ is a variable and $e_u, e_v$ are two expressions with values in $\ZZ$, while $e_{\operatorname{\textit{in}}}$ and $e_{\operatorname{\textit{out}}}$ are boolean expressions with values in $\{0,1\}$. Our construct generalizes `$\FF{for}\, r \, \FF{\{}P\FF{\}}$' in \MSRL (Section \ref{section:forest is complete and sound with respect to MSRL} will recall \MSRL) by restricting \JANUS \cite{LUTZ:JANUS86} iteration, which let \JANUS be (reversible) Turing-complete. Construct~\eqref{align:from-to forest introduction} assures that \forest iterations simultaneously enjoy the two following features: (i) they can be halted by means of predicates, providing more control over the computation flow; (ii) \forest is compatible with \textit{structured programming}, possibly easing formal reasoning on it \cite{DBLP:books/mc/22/Dijkstra22e}.

\subsubsection*{Iteration in \forest, intuitively.}
We conclude this introduction by illustrating how~\eqref{align:from-to forest introduction} restricts the iteration in \JANUS.
The body $P$ of~\eqref{align:from-to forest introduction} cannot alter the variable $i$ which drives the iteration. Entering the iteration is under the control of a logical disjunction with form $i\FF{=}e_u\,\FF{or}\,e_{\operatorname{\textit{in}}}$. Analogously, exiting the iteration is under the control of a logical disjunction with form $i\FF{=}e_v\,\FF{or}\,e_{\operatorname{\textit{out}}}$.
\par
Assuming that $e_u$ evaluates to $u$, and $e_v$ to $v$ such that $u \leq v$, an iteration starts looping if $i$ belongs to the interval $[u,v]$ with the proviso that, in case $i$ is strictly greater than $u$, then $e_{\operatorname{\textit{in}}}$ must be true. Under the initial assumption, every iteration increments $i$ by one unit.
So, the iteration keeps going until
$i\FF{=}e_v\,\FF{or}\,e_{\operatorname{\textit{out}}}$
holds true, namely until $i$ reaches the upper bound $v$, or the exit condition $e_{\operatorname{\textit{out}}}$ becomes true.
\begin{figure}
	\centering
	\begin{tikzpicture}
		\fill[fill = lightgray] (-0.3,0) rectangle (8.3,1);
		\fill[fill = GreenYellow] (1,0) rectangle (7,.5);
		\draw[draw = none] (1,0) rectangle node[text = black]{} (7,.5);

		\path
			(-0.3,0)   node(yDot)[shape=circle][fill, inner sep = 1.25pt]{}
			(-0.35,-0.2)   node(y)[shape=circle] {$e_u$}

			(1,0.5) node(iStartDot)[circle,fill, inner sep=1.25pt]{}
			(0.8,1)  node(iStart)[shape=circle] 	{}
			(1,1.3) node(iEnd)[shape = circle]{}
			(1,0) node(iStartBottom)[shape=circle][circle,fill, inner sep=1.25pt]{}
			(1.7,-0.2)	node(w)[shape = circle] {$e_u \FF{<}i \ \FF{and} \ e_{\operatorname{\textit{in}}}$}

			(7,.5) node(iEndDot)[circle,fill,inner sep = 1.25pt]{}
			(7,1.3) node(iEnd)[shape=circle]{}
			(7,0) node(iEndBottom)[shape=circle][circle,fill, inner sep=1.25pt]{}
			(6.3,-0.2)	node(k)[shape = circle] {$i \FF{<} e_v\ \FF{and} \ \FF{!}\!e_{\operatorname{\textit{out}}}$}

			(8.3,0)   node(xDot)[shape=circle][circle,fill, inner sep=1.25pt]{}
			(8.35,-0.2)  node(x)[shape=circle] 	{$e_v$};

		\draw[-{Latex}] (-1,0) -- (9,0);
		\draw[-{Latex}] (iStartDot) -- (iEndDot);
		\draw[thick,dash dot] (iStartBottom) -- (iStartDot);
		\draw[thick,dash dot] (iEndDot) -- (iEndBottom);
		\draw[dash dot] (yDot) -- (-0.3,1);
		\draw[dash dot] (xDot) -- (8.3,1);
		\draw[dash dot] (-0.3,1) -- (8.3,1);
	\end{tikzpicture}
\caption{The iteration of \forest begins/halts depending on $e_u, e_v, e_{\operatorname{\textit{in}}}$, and $e_{\operatorname{\textit{out}}}$.}
\label{figure:iteration early start stop}
\end{figure}
\noindent
Figure~\ref{figure:iteration early start stop} visually summarizes how $i$ moves inside the interval $[u,v]$, highlighting that the difference $v-u$ sets the maximum amount of iterations.

On the other side, if $u > v$ when the iteration starts, $i$ must belong to the interval $[v,u]$, and the loop develops a computation which is the inverse of the one we have just described. The flow-charts in Figure~\ref{fig:LoopExplanation}, which we specialized from \cite{YOKOYAMAAXELSENGLUCK:TCS16}, will fully describe the computational flow of~\eqref{align:from-to forest introduction}.

\subsubsection*{Structure of the work.}
Section~\ref{section:The language Forest} introduces syntax and operational semantics of \forest.
Section~\ref{section:Properties of Forest} firstly shows that the operational semantics always terminates when interpreting a term $P$ in \forest, even though this does not mean that $P$ always produces a meaningful state. Secondly, it shows that the function in Section~\ref{section:The language Forest}, which defines $P^-$ for any $P$ in \forest, actually yields the reverse of $P$.
Section~\ref{section:forest is complete and sound with respect to MSRL} translates \MSRL into \forest, proving that the latter is complete with respect to the first one.
Section~\ref{section:Algoritmic expressivity: SRL vs forest} shows that \forest is algorithmically more expressive than \MSRL. Section~\ref{section:Conclusions, future work} concludes, pointing to future and related work.

\section{The computational model \forest}
\label{section:The language Forest}
Concerning the algorithmic expressiveness, \forest is defined by means of a syntax and of an operational semantics, designed to fall in between \MSRL and \JANUS.

\subsubsection*{Syntax.}
After some preliminaries (Definition~\ref{definition:Set of expressions}) the structure of every term $P$ is given by simultaneously defining \emph{domain} and \emph{writable domain} of $P$ to assure that \forest contains the inverse of $P$ itself (Definition~\ref{definition:Set forestP well-formed terms}).

\begin{definition}[Arithmetical and boolean expressions]
\label{definition:Set of expressions}
\begin{enumerate}
\item
Let $V$ be a set of variable names \FF{x}, \FF{y}, \ldots.
Let $\EXPRZ$ denote the set of arithmetical expressions. Representation of numbers in $\ZZ$, and elements of $V$ belong to $\EXPRZ$. Moreover, if $e, e'\in\EXPRZ$, then their \emph{sum} and \emph{subtraction} is in $\EXPRZ$. Finally, the \emph{domain} $\DOM{e}$ of $e \in\EXPRZ$ is the set of variables of $V$ occurring in $e$.
\item
Let $\EXPRB$ denote the set of boolean expressions. Truth values $\FF{0}, \FF{1}$, and the \emph{equivalence test} $e\FF{=}e'$ are in $\EXPRB$, for every $e, e'\in\EXPRZ$.
Moreover, $\EXPRB$ contains \emph{disjunctions}, \emph{conjunctions} and \emph{negations} (operator `\FF{!}') of elements already in $\EXPRB$. Finally, $\DOM{e}$ is the \emph{domain} of $e\in\EXPRB$ with all the variables of $V$ in $e$.
\end{enumerate}
\end{definition}
\noindent
For example, $\FF{(-1)+(x-3)-y} \in \EXPRZ$ and $\DOM{\FF{(-1)+(x-3)-y}}=\{\FF{x},\FF{y}\} $ while $\FF{(3=y)}$ $\FF{or!(1=x+y)} \in \EXPRB$, and $\DOM{\FF{(3=y)or!(1=x+y)}}$ $= \{\FF{x},\FF{y}\}$.

\begin{definition}[Set $\forestP$ of \emph{well-formed terms}]
\label{definition:Set forestP well-formed terms}
The elements of $\forestP$ are ranged over by $P, Q$ \ldots, and are defined together with their \emph{domain} $\DOM{P}\subseteq V$ and \emph{writable domain} $\WDOM{P} \subseteq \DOM{P}$ as follows:
\begin{enumerate}
    \item
    $\FF{skip}\in\forestP$
    with $\DOM{\FF{skip}} = \WDOM{\FF{skip}} = \emptyset$;

    \item
    \label{definition:Set forestP well-formed terms point 2}
    let $x\in V$, and $e\in \EXPRZ$ s.t. $x \notin \DOM{e}$.
    The \emph{assignments} $x\FF{+=}e, x\FF{-=}e\in\forestP$
    with $\DOM{x\FF{+=}e} = \DOM{x\FF{-=}e} = \DOM{e} \cup \{x\}$,
    and $\WDOM{x\FF{+=}e} = \WDOM{x\FF{-=}e} = \{x\}$;

    \item
	let $P, Q \in \forestP$. The \emph{series composition} $P\FF{;}Q\in\forestP$
    with $\DOM{P\FF{;}Q} = \DOM{P}\cup\DOM{Q}$,
    and $\WDOM{P\FF{;}Q} = \WDOM{P}\cup\WDOM{Q}$;

    \item
	let $e \in \EXPRB$ and $P,Q \in \forestP$ such that $\DOM{e} \cap (\WDOM{P} \cup \WDOM{Q}) = \emptyset$. The \emph{selection}
    $\FF{if(}e\FF{)\{}P\FF{\}else\{}Q\FF{\}} \in \forestP$
	with \emph{guard} $e$, \emph{domain} $\DOM{e}\cup\DOM{P}\cup\DOM{Q}$,
    and \emph{writable domain} $\WDOM{P} \cup \WDOM{Q}$;

	\item
    let $i\in V$, $e_u, e_v \in \EXPRZ$,
	and $e_{\operatorname{\textit{in}}}, e_{\operatorname{\textit{out}}} \in \EXPRB$.
	Let $P \in \forestP$ such that $(\{i\} \cup \DOM{e_u} \cup \DOM{e_v})\cap \WDOM{P} = \emptyset$.
	The \emph{iteration}:
	\begin{align}
    \label{aligh:from-to definition}
		& \FF{from (}i\FF{=}e_u\,\FF{or}\,e_{\operatorname{\textit{in}}}\FF{) to (}i\FF{=}e_v\,\FF{or}\,e_{\operatorname{\textit{out}}}\FF{)\{} P \FF{\}}
	\end{align}
    belongs to $\forestP$
	with \emph{domain} $\DOM{e_{\operatorname{\textit{in}}}}\cup\DOM{e_{\operatorname{\textit{out}}}}\cup\DOM{e_u}\cup\DOM{e_v}\cup\DOM{P} \cup \{i\}$ and \emph{writable domain} $\WDOM{P} \cup \{i\}$.
    We will typically refer to \eqref{aligh:from-to definition} as ``\FF{from-to} term'' and to $i$ as \emph{iteration} or \emph{leading variable}.
\end{enumerate}
\end{definition}
\noindent
Listing~\ref{listing:Minimum minimumF in forest} gives an example of a term in $\forestP$.
Lines 4--7 have \emph{domain} \FF{x}, \FF{y}, \FF{min}, \FF{found} and \emph{writable domain} \FF{min}, \FF{found}. Neither \FF{min}, nor \FF{found} occurs in the domain of \FF{(i=y)}, \FF{(i=0)}, and \FF{(i=x)} at lines 3 and 2, respectively.
This is true in general. For any $P\in\forestP$, the notion ``\emph{writable domain}'' implies that $P$ cannot write into variables that belong to the domain of the guard of any selection, or to the leading variable or variables in domains of the bounds $e_u$, $e_v$ of any iteration having $P$ as a sub-term.

\begin{definition}[Inverse of a term]
\label{definition:Inverse of a term}
The inverse $P^-$ of $P\in\forestP$ is defined inductively.
The inverse of \FF{skip} is \FF{skip}.
The inverse of $x\FF{+=}e$, and $x\FF{-=}e$, are $x\FF{-=}e$, and $x\FF{+=}e$, respectively. Let $P,Q \in \forestP$. The inverse of $P\FF{;}Q$  is $Q^-\FF{;}P^-$,
while $\FF{if(}e\FF{)\{}P^-\FF{\}else\{}Q^-\FF{\}}$ is the
inverse of $\FF{if(}e\FF{)\{}P\FF{\}then\{}Q\FF{\}}$. Finally:
\begin{align}
\label{align:inverse of from-to}
\left(\begin{array}{ll}
	\FF{from(}i \FF{=} e_u\ \FF{or}\ e_{\operatorname{\textit{in}}}\FF{)} \\
	\quad \FF{to(}i \FF{=} e_v\ \FF{or}\ e_{\operatorname{\textit{out}}}\FF{)\{}P\FF{\}}
\end{array}\right)^{-}
= \begin{array}{ll}
	\FF{from(}i \FF{=} e_v\ \FF{or}\ e_{\operatorname{\textit{out}}}\FF{)} \\
	\quad\FF{to(}i \FF{=} e_u\ \FF{or}\ e_{\operatorname{\textit{in}}}\FF{)\{}P\FF{\}}
	\enspace .
\end{array}
\end{align}
\end{definition}
\par
It should be obvious that the image of the here above function is $\forestP$. Moreover, the rightmost $P$ in \eqref{align:inverse of from-to} \emph{is not} a typo. We choose the form of \eqref{align:inverse of from-to} between two alternatives offered by generalizing Matos' \MSRL iteration to \FF{from-to}. Our choice is to encode the inversion by exchanging lower limit $e_u$, and upper limit $e_v$ whose values set the range for $i$, and to design an operational semantics able to let
$\FF{from(}i\FF{=}e_v\ \FF{or}\ e_{\operatorname{\textit{out}}}\FF{)to(}i\FF{=}e_u\ \FF{or}$ $e_{\operatorname{\textit{in}}}\FF{)\{}P\FF{\}}$ be the inverse of
$\FF{from(}i\FF{=}e_u\ \FF{or}\ e_{\operatorname{\textit{in}}}\FF{)to}$ $\FF{(}i\FF{=}e_v \ \FF{or}\ e_{\operatorname{\textit{out}}}\FF{)\{}P\FF{\}}$.
An alternative would be to include a further iterative construct, name it
$\FF{morf(}i\FF{=}e_v\ \FF{or}\ e_{\operatorname{\textit{out}}}\FF{)ot(}i\FF{=}e_u\, \FF{or}\ e_{\operatorname{\textit{in}}}\FF{)\{}P\FF{\}}$ for example, such that:
\begin{align}
\label{align:inverse of from-to alternative}
\left(\begin{array}{ll}
    \FF{from(}i\FF{=}e_u\ \FF{or}\ e_{\operatorname{\textit{in}}}\FF{)} \\
    \quad \FF{to(}i\FF{=}e_v\ \FF{or}\ e_{\operatorname{\textit{out}}}\FF{)\{}P\FF{\}}
\end{array}\right)^{-}
&= \begin{array}{ll}
    \FF{morf(}i\FF{=}e_u\ \FF{or}\  e_{\operatorname{\textit{out}}}\FF{)} \\
    \quad\FF{ot(}i\FF{=}e_v\ \FF{or}\ e_{\operatorname{\textit{in}}}\FF{)\{}P^-\FF{\}}
    \enspace ,
\end{array}
\end{align}
\noindent
which would be a choice more coherent than~\eqref{align:inverse of from-to} as compared to how Matos actually inverts iteration in \MSRL \cite{MATOS:TCS15}. We opted for syntactic compactness.
Our choice would correspond to define \MSRL with an iteration $\FF{for}\ r\ \FF{\{}Q\FF{\}}$ whose inverse is $\FF{for}\ \FF{-}r\ \FF{\{}Q\FF{\}}$, such that, if $r < 0$, then $Q^-$ is iterated $|r|$ times.

\subsubsection*{Operational semantics.}
A state (Definition~\ref{definition:States}) sets the values of variables, allowing us to evaluate expressions (Definition~\ref{definition:Evaluating arithmetic and boolean expressions}, and Figure~\ref{fig:interpretation of boolean expressions}). Figure~\ref{fig:operationalSemantics} introduces the rules to interpret terms that belong to a syntax which extends $\forestP$ given in Definition~\ref{definition:Set forestP well-formed terms}. \emph{This is necessary to formalize the behavior of a reversible computational model by means of a classical one.}

\begin{definition}[States]
\label{definition:States}
A \textit{state} is a \emph{total} map $\bsigma: V \rightarrow \mathbb{Z}$.
We write $\bsigma[x_1\! \mapsto\! v_1 \ldots x_n\! \mapsto\! v_n]$ to denote a state such that
$\bsigma[x_1\! \mapsto\! v_1 \ldots x_n\!\mapsto\! v_n](y) = v_i $,
if $y = x_i$, for any $i\in\{1, \ldots, n\}$; otherwise
$\bsigma[x_1\! \mapsto\! v_1 \ldots x_n\! \mapsto\! v_n](y) = \bsigma(y)$.
The set $\Sigma$, ranged over by $\bsigma, \bsigma' \ldots \btau, \btau' \ldots$, contains all the possible states.
\end{definition}

\begin{figure}
    \begin{align*}
        \bsigma, \FF{0} & \Downarrow 0
        &
        \bsigma, \FF{1} & \Downarrow 1
        &
        \bsigma, e \FF{=} e' & \Downarrow
        \begin{cases}
            1 & (\bsigma, e \Downarrow n) \wedge (\bsigma, e' \Downarrow n) \\
            0 & \textrm{otherwise}
        \end{cases}
    \end{align*}
    \begin{align}
        \label{align:boolean expression interpretation not}
        \bsigma, \FF{!}e & \Downarrow 1-m
        && (\bsigma, e \Downarrow m)
        \\
        \label{align:boolean expression interpretation and}
        \bsigma, e\, \FF{and}\, e' & \Downarrow m \cdot n
        && (\bsigma, e \Downarrow m) \wedge (\bsigma, e' \Downarrow n)
        \\
        \label{align:boolean expression interpretation or}
        \bsigma, e\, \FF{or}\, e' & \Downarrow m+n-m\cdot n
        && (\bsigma, e \Downarrow m) \wedge (\bsigma, e' \Downarrow n)
    \end{align}
    \caption{Interpretation of boolean expressions.}
    \label{fig:interpretation of boolean expressions}
\end{figure}

\begin{definition}[Evaluating arithmetic and boolean expressions]
\label{definition:Evaluating arithmetic and boolean expressions}
Let $e \in \EXPRZ \cup \EXPRB$. 
We write $\bsigma, e \Downarrow n$ meaning that $e$ \textit{evaluates} to $n$ in the state $\bsigma$.
If $e \in \EXPRZ$, the relation $\Downarrow$ can be defined obviously on the structure of $e$.
Concerning boolean expressions, the meaning of $\bsigma, e \Downarrow n$ is as in {\normalfont Figure~{\ref{fig:interpretation of boolean expressions}}}.
\end{definition}

\begin{remark}
Clauses~\eqref{align:boolean expression interpretation not}, \eqref{align:boolean expression interpretation and}, and \eqref{align:boolean expression interpretation or} interpret boolean expressions by quadratic polynomials, yielding $\{0, 1\}$ iff $\bsigma$ assigns values $\{0, 1\}$ to every variable of $\FF{!}e$, $e\, \FF{and}\, e'$, $e\, \FF{or}\, e'$. \qed
\end{remark}

\begin{definition}[Extension $\forestPext$ of $\forestP$]
\label{definition:Extension forestPext of forestP}
Let $\forestPext$ be $\forestP$ with boolean expressions $\EXPRB$ that also contain `$e_u\FF{<=}e_v$' and `$e_u\FF{>}e_v$', with $e_u,e_v \in \EXPRB$, and extended to contain new terms:
\begin{align*}
    &\FF{assert(}e\FF{)}
    &
    &\FF{loop until(}e\ \FF{or}\ e_{\operatorname{\textit{out}}}\FF{)\{}P{\}}
    \enspace .
\end{align*}
\end{definition}
\noindent
For example, \FF{if(u<=v)\{assert(x<1);skip\}else\{loop until(z>2)\{skip\}\}} belongs to $\forestPext$.

\begin{figure}
	\centering
	\begin{tabular}{c}
		\bottomAlignProof
		\AxiomC{$\bsigma, x \Downarrow n$}
		\AxiomC{$\bsigma, e \Downarrow m$}
		\RightLabel{\scriptsize \textsc{Inc}}
		\BinaryInfC{$\bsigma \ \, x\FF{+=}e \ \, \bsigma[x \mapsto n + m]$}
		\DisplayProof
		\qquad
		\bottomAlignProof
		\AxiomC{$\bsigma, x \Downarrow n$}
		\AxiomC{$\bsigma, e \Downarrow m$}
		\RightLabel{\scriptsize \textsc{Dec}}
		\BinaryInfC{$\bsigma \ \, x\FF{-=}e \ \, \bsigma[x \mapsto n - m]$}
		\DisplayProof
		\\ \\
		\bottomAlignProof
		\AxiomC{$\bsigma, b \Downarrow 1$}
		\AxiomC{$\bsigma \ \, P \ \, \bsigma'$}
		\RightLabel{\scriptsize \textsc{IfTrue}}
		\BinaryInfC{$\bsigma \ \, \FF{if(}b\FF{)\{}P\FF{\}else\{}Q\FF{\}} \ \, \bsigma'$}
		\DisplayProof
		\quad
		\bottomAlignProof
		\AxiomC{$\bsigma, b \Downarrow 0$}
		\AxiomC{$\bsigma \ \, Q \ \, \bsigma'$}
		\RightLabel{\scriptsize \textsc{IfFalse}}
		\BinaryInfC{$\bsigma \ \, \FF{if(}b\FF{)\{}P\FF{\}else\{}Q\FF{\}} \  \, \bsigma'$}
		\DisplayProof
		\\ \\
\bottomAlignProof
\AxiomC{}
\RightLabel{\scriptsize \textsc{Skip}}
\UnaryInfC{$\bsigma \ \, \FF{skip} \ \, \bsigma$}
\DisplayProof
\qquad
\bottomAlignProof
\AxiomC{$\bsigma\ \,P\ \,\bsigma'$}
\AxiomC{$\bsigma'\ \,Q\ \,\bsigma''$}
\RightLabel{\scriptsize \textsc{Seq}}
\BinaryInfC{$\bsigma\ \,P\FF{;}Q\ \,\bsigma''$}
\DisplayProof
\\ \\
\bottomAlignProof
\AxiomC{\hspace{-15pt}
	$\begin{matrix}
		\bsigma\ \,\FF{if(}e_u\FF{<=}e_v
              \FF{)\{assert(}e_u\FF{<=}i\ \FF{and}\ i\FF{<=}e_v\FF{);assert(}i\FF{=}e_u\ \FF{or}\ e_{\operatorname{\textit{in}}}\FF{);}
              \\\qquad\qquad\qquad\qquad\
		\FF{loop until(}i\FF{=}e_v\ \FF{or}\ e_{\operatorname{\textit{out}}}\FF{)\{}
		P\FF{;}i\FF{+=1}\FF{;assert(!}e_{\operatorname{\textit{in}}}\FF{)\}}
        \\
		\FF{\}else\{assert(}e_v\FF{<=}i\ \FF{and}\ i\FF{<=}e_u\FF{);assert(}i\FF{=}e_u\ \FF{or}\ e_{\operatorname{\textit{in}}}\FF{);}
              \\\qquad\qquad\qquad\qquad\
		\FF{loop until(}i\FF{=}e_v\ \FF{or}\ e_{\operatorname{\textit{out}}}\FF{)\{}
		i\FF{-=1}\FF{;}P^-\FF{;assert(!}e_{\operatorname{\textit{in}}}\FF{)}
		\FF{\}\}}
		\ \, \btau
	\end{matrix}$}
\RightLabel{\scriptsize \textsc{FromTo}}
\UnaryInfC{$\bsigma\ \,
\FF{from(}i\FF{=}e_u\ \FF{or}\ e_{in}\FF{)to(}i\FF{=}e_v\ \FF{or}\ e_{\operatorname{\textit{out}}}\FF{)\{}
P\FF{\}}\ \,\btau$
}
\DisplayProof
\\    \\
\bottomAlignProof
\AxiomC{$\bsigma,\, e \Downarrow 1$}
\RightLabel{\scriptsize \textsc{Assert1}}
\UnaryInfC{$\bsigma \ \,
    \FF{assert(}e\FF{)} \ \,
    \bsigma$
}
\DisplayProof
\quad\
\bottomAlignProof
\AxiomC{$\bsigma,\, e \Downarrow 0$}
\RightLabel{\scriptsize \textsc{Assert0}}
\UnaryInfC{$\bsigma \ \,
    \FF{assert(}e\FF{)} \ \,
    \bbot$
}
\DisplayProof
\quad\
\bottomAlignProof
\AxiomC{}
\RightLabel{\scriptsize \textsc{Prop}}
\UnaryInfC{$\bbot \ \, P \ \,\bbot$
}
\DisplayProof
\\ \\
\bottomAlignProof
\AxiomC{$\bsigma, \FF{(}i\FF{=}e_v\ \FF{or}\ e_{\operatorname{\textit{out}}}\FF{)} \Downarrow 1$}
\AxiomC{$\bsigma\ \,\FF{skip}\ \,\btau$}
\RightLabel{\scriptsize \textsc{LoopBase}}
\BinaryInfC{$\bsigma \ \,
    \FF{loop until(}i\FF{=}e_v\ \FF{or}\ e_{\operatorname{\textit{out}}}\FF{)\{}\,
    P\,
    \FF{\}} \ \, \btau$
}
\DisplayProof
\\ \\
\AxiomC{
$\bsigma, \FF{(}i\FF{=}e_v\ \FF{or}\ e_{\operatorname{\textit{out}}}\FF{)} \Downarrow 0 $
}
\AxiomC{
    $\bsigma \ \,
    P
    \FF{;loop until(}i\FF{=}e_v\ \FF{or}\ e_{\operatorname{\textit{out}}}\FF{)\{}\,
    P\,
    \FF{\}} \ \,
    \btau$
}
\RightLabel{\scriptsize \textsc{LoopRec}}
\BinaryInfC{$\bsigma \ \,
    \FF{loop until(}i\FF{=}e_u\ \FF{or}\ e_{\operatorname{\textit{out}}}\FF{)\{}\,
    P\,
    \FF{\}} \ \,
    \btau$
}
\DisplayProof
	\end{tabular}
	\caption{Operational Semantics on $\forestPext$.}
	\label{fig:operationalSemantics}
\end{figure}

\begin{definition}[Operational semantics of $\forestPext$]
\label{definition:Operational semantics for an extension of F}
{\normalfont Figure~{\ref{fig:operationalSemantics}}} introduces the rules to interpret terms of $\forestPext$. The rules derive three kinds of judgments: $\bsigma\,P\,\btau$, $\bsigma\,P\,\bbot$, and $\bbot\,P\,\bbot$, for $\bsigma$, $\btau$ in $\Sigma$, where $\bbot$, which is not in $\Sigma$, denotes the result of a failing interpretation.
\par
\emph{Interpreting $P$} means to fix a state $\bsigma$
and to build a derivation tree with rules in {\normalfont Figure~{\ref{fig:operationalSemantics}}} whose conclusion can be either a \emph{successful} $\bsigma\,P\,\btau$, for some $\btau\in\Sigma$, or a \emph{failing} $\bsigma\,P\,\bbot$, the latter generated if one judgment between $\bmu\ \FF{assert(}e\FF{)} \, \bbot$ or $\bbot\,Q\,\bbot$ shows up while building the tree, for some $\bmu, e$, and $Q$.
\end{definition}
\noindent
Rules \textsc{Inc}, \textsc{Dec}, \textsc{IfFalse}, \textsc{IfTrue}, and \textsc{Skip} interpret terms in $\forestPext$ as expected.
\textsc{Seq} decomposes the evaluation of a sequence $P\FF{;}Q$, to produce $\bsigma'$ required by $Q$ the application of \textsc{Seq} proceeds deterministically from left to right.
    \begin{figure}
	\begin{subfigure}{\textwidth}
		\vspace{0.5\baselineskip}
		\centering
		\begin{tikzpicture}[scale=1]
			\coordinate (gLoopFalse) at (-1, -0.15);
			\coordinate (giTrue) at (-2.25,1);

			\node[shape = circle] (start) at (-5.5,-0.15) {$\bsigma$};
			\node[shape = circle, draw] (giInt) at (-4, -0.15)
			{\scalebox{.5}{%
					$\begin{array}{c}
						e_u\FF{<=}i\\
						\FF{and}\\
						i\FF{<=}e_v
					\end{array}$}};
			\node[shape=circle,draw] (gi) at (-2.25,-0.15)
			{\scalebox{.5}{%
					$\begin{array}{c}
						i\FF{=}e_u\\
						\FF{or} \\
						e_{\operatorname{\textit{in}}}
					\end{array}$}};

			\node[shape=circle,draw] (gLoop) at (0.5,-0.15)
			{\scalebox{.5}{%
					$\begin{array}{c}
						\FF{!}e_{\operatorname{\textit{in}}}
					\end{array}$}
			};

			\node (ground) at (-2.25,-1.5) {$\bbot$};

			\node[shape=rectangle,draw] (gl) at (0.5,-1.25) {\scalebox{.5}{$i\FF{+=1}$}};
			\node[shape=diamond,draw] (d) at (3,-0.15)
			{\scalebox{.5}{
					$\begin{array}{c}
						i\FF{=}e_v \\
						\FF{or}\\
						e_{\operatorname{\textit{out}}}
					\end{array}$}};
			\node[shape=rectangle,draw] (P') at (2,-1.25) {$P$};
			\node[shape = circle] (stop) at (4.745,-0.15) {$\btau$};
			\coordinate (stop') at (4.5,-0.15);

			\draw[->] (giInt) -- node[above]{\scriptsize \Circled{1}} (gi);
			\draw[->] (start) -- node[above]{\scriptsize \Circled{0}} (giInt);
			\draw[-]  (gi) --  (giTrue);
			\draw[->] (giTrue) -| (d);
			\draw[->] (d) -- node[above]{\scriptsize true} (stop');
			\draw[->] (d) |- node[below]{\scriptsize false} (P');
			\draw[->] (P') --  node[below]{\scriptsize \Circled{2}} (gl);
			\draw[->] (gi) -- node[near start, right]{\scriptsize false} (ground);
			\draw[->] (giInt) |- node[near start, right]{\scriptsize false} (ground);
			\draw[->] (gl) --  node[left]{\scriptsize \Circled{3}} (gLoop);
			\draw[->] (gLoop) -- (d);
			\draw[-] (gLoop) -- node[near start, above]{\scriptsize false} (gLoopFalse);
			\draw[->] (gLoopFalse) |- (ground);
		\end{tikzpicture}
		\caption{Flow-chart if $e_u\FF{<=}e_v$.}
		\label{fig: Computational flow of from-to forward}
	\end{subfigure}

	\begin{subfigure}{\textwidth}
		\centering
		\vspace{\baselineskip}
		\begin{tikzpicture}[scale=1]
			\coordinate (gLoopFalse) at (-1, -0.15);
			\coordinate (giTrue) at (-2.25,1);

			\node[shape = circle] (start) at (-5.5,-0.15) {$\bsigma$};
			\node[shape = circle, draw] (giInt) at (-4, -0.15)
			{\scalebox{.5}{%
					$\begin{array}{c}
						e_v\FF{<=}i\\
						\FF{and}\\
						i\FF{<=}e_u
					\end{array}$}};
			\node[shape=circle,draw] (gi) at (-2.25,-0.15)
			{\scalebox{.5}{%
					$\begin{array}{c}
						i\FF{=}e_u\\
						\FF{or} \\
						e_{\operatorname{\textit{in}}}
					\end{array}$}};

			\node[shape=circle,draw] (gLoop) at (0.5,-0.15)
			{\scalebox{.5}{%
					$\begin{array}{c}
						\FF{!}e_{in}
					\end{array}$}
			};

			\node (ground) at (-2.25,-1.5) {$\bbot$};

			\node[shape=rectangle,draw] (gl) at (0.5,-1.25) {$P^-$};
			\node[shape=diamond,draw] (d) at (3,-0.15)
			{\scalebox{.5}{
					$\begin{array}{c}
						i\FF{=}e_v \\
						\FF{or}\\
						e_{\operatorname{\textit{out}}}
					\end{array}$}};
			\node[shape=rectangle,draw] (P') at (2,-1.25) {\scalebox{.5}{$i\FF{-=1}$}};
			\node[shape = circle] (stop) at (4.745,-0.15) {$\btau$};
            \coordinate (stop') at (4.5,-0.15);

\draw[->] (giInt) -- node[above]{\scriptsize \Circled{1}} (gi);
\draw[->] (start) -- node[above]{\scriptsize \Circled{0}} (giInt);
\draw[-]  (gi) --  (giTrue);
\draw[->] (giTrue) -| (d);
\draw[->] (d) -- node[above]{\scriptsize true} (stop');
\draw[->] (d) |- node[below]{\scriptsize false} (P');
\draw[->] (P') --  node[below]{\scriptsize \Circled{2}} (gl);
\draw[->] (gi) -- node[near start, right]{\scriptsize false} (ground);
\draw[->] (giInt) |- node[near start, right]{\scriptsize false} (ground);
\draw[->] (gl) --  node[left]{\scriptsize \Circled{3}} (gLoop);
\draw[->] (gLoop) -- (d);
\draw[-] (gLoop) -- node[near start, above]{\scriptsize false} (gLoopFalse);
\draw[->] (gLoopFalse) |- (ground);

		\end{tikzpicture}
		\caption{Flow-chart if $e_u\FF{>}e_v$.}
		\label{fig: Computational flow of from-to backward}
	\end{subfigure}
	\caption{Flow-chart of the premise of rule \textsc{FromTo} in Figure~\ref{fig:operationalSemantics}.}
	\label{fig:LoopExplanation}
\end{figure}
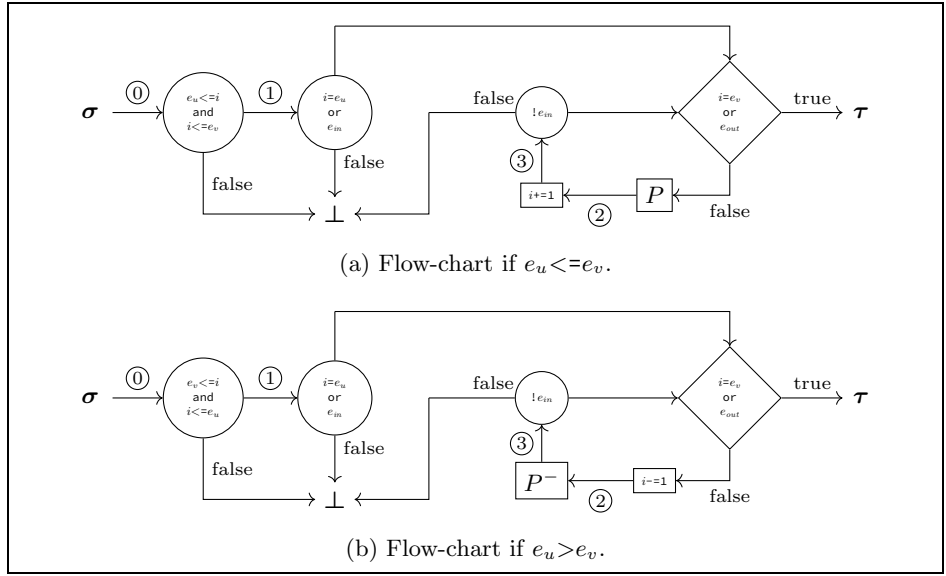
\noindent
\textsc{FromTo} interprets the eponymous term of $\forestPext$. Being $\bsigma$ an initial state, the premise of \textsc{FromTo} triggers either \textsc{IfTrue} or \textsc{IfFalse} to interpret \emph{exactly} one between the sequences:
\begin{equation}
\begin{aligned}
\label{aligned:loop-until branch true}
&\Circled{0}\,\FF{assert(}e_u\FF{<=}i\ \FF{and}\ i\FF{<=}e_v\FF{);}
 \Circled{1}\,\FF{assert(}i\FF{=}e_u\ \FF{or}\ e_{\operatorname{\textit{in}}}\FF{);}
\\
&\FF{loop until(}i\FF{=}e_v\ \FF{or}\ e_{\operatorname{\textit{out}}}\FF{)\{}P\FF{;}
 \Circled{2}\,i\FF{+=1;}
 \Circled{3}\,\FF{assert(!}e_{\operatorname{\textit{in}}}\FF{)\}}
\end{aligned}
\end{equation}
\begin{equation}
\begin{aligned}
\label{aligned:loop-until branch false}
&\Circled{0}\,\FF{assert(}e_v\FF{<=}i\ \FF{and}\ i\FF{<=}e_u\FF{);}
 \Circled{1}\,\FF{assert(}i\FF{=}e_u\ \FF{or}\ e_{\operatorname{\textit{in}}}\FF{);}\\
&\FF{loop until(}i\FF{=}e_v\ \FF{or}\ e_{\operatorname{\textit{out}}}\FF{)\{}i\FF{-=1;}
 \Circled{2}\,P^-\FF{;}
 \Circled{3}\,\FF{assert(!}e_{\operatorname{\textit{in}}}\FF{)\}}
\end{aligned}
\end{equation}
\noindent
by means of \textsc{LoopRec}, \textsc{LoopBase}, \textsc{Assert1}, and \textsc{Assert0}. Numbers \Circled{0}, \Circled{1} {\ldots\xspace} in \eqref{aligned:loop-until branch true} trace how \eqref{aligned:loop-until branch true} corresponds to the flow-chart in Figure~\ref{fig: Computational flow of from-to forward}, which is interpreted if $\bsigma, e_u\FF{<=}e_v \Downarrow 1$, i.e. if the argument of the selection in the premise of \textsc{FromTo}, holds true. Initially, the interpretation of \eqref{aligned:loop-until branch true} can yield $\bbot$ in two cases because of the first two \FF{assert}. The first \FF{assert} fails if $\bsigma, e_u \FF{<=} i \, \FF{and} \, i \FF{<=} e_v \Downarrow 0$, i.e. the value of $i$ is not in $[u,v]$, assuming that $e_u, e_v$ evaluate to $u, v$, respectively. Being the first \FF{assert} successful, the second one fails if $\bsigma, i\FF{=}e_u\ \FF{or}\ e_{\operatorname{\textit{in}}} \Downarrow 0$, i.e.  the value of $i$ is greater than $u$ but the value of $e_{\operatorname{\textit{in}}}$ forbids to enter the loop, namely it holds false. Otherwise, the construct \FF{loop-until} starts and may keep unfolding until neither $i$ reaches $v$, nor $e_{\operatorname{\textit{out}}}$ decrees to exit, namely it holds true. Rules \textsc{LoopRec}, and \textsc{LoopBase} govern the unfolding, generating:
\begin{equation}
    \begin{aligned}
        \label{aligned:loop-until unfolding}
        &P\FF{;}\Circled{2}\,i\FF{+=1;}\Circled{3}\,\FF{assert(!}e_{\operatorname{\textit{in}}}\FF{);}
         P\FF{;}\Circled{2}\,i\FF{+=1;}\Circled{3}\,\FF{assert(!}e_{\operatorname{\textit{in}}}\FF{);}
         \ldots
    \end{aligned}
\end{equation}
\noindent
which may be (abnormally) stopped, yielding $\bbot$, as soon as $P$ sets $\FF{!}e_{\operatorname{\textit{in}}}$ to false. If this happens, $\bbot$ is propagated by means of \textsc{Prop}, leading to the failure of the whole interpretation. If the argument $e_u\FF{<=}e_v$ of the selection in the premise of \textsc{FromTo} holds false in the initial state $\bsigma$, then \eqref{aligned:loop-until branch false} must be interpreted, according the flow-chart in Figure~\ref{fig: Computational flow of from-to backward}, whose behavior is analogous to the one in Figure~\ref{fig: Computational flow of from-to forward}.

\begin{definition}[The computational model \forest]
\label{definition:The language forest}
\forest has $\forestP$ in {\normalfont Definition~\ref{definition:Set forestP well-formed terms}} as its syntax, and operational semantics as in {\normalfont Definition~\ref{definition:Operational semantics for an extension of F}}. By $P\in\forest$ we mean $P\in\forestP$.
\end{definition}

\begin{remark}[\forest syntax and operational semantics]
\label{remark:On the syntax and the operational semantics of forest}
\begin{enumerate}

\item Definition~\ref{definition:The language forest} says that a term in \forest cannot be written using the syntax in $\forestPext$: the terms of $\forestPext$ are \emph{invisible} to a ``programmer''.

\item The terms of $\forest$ are designed to be \emph{interpreted} deterministically.
Given any $P\in\forestP$, two cases exist.
On one side, $P$ can be interpreted from $\bot$: in this case only \textsc{Prop} applies, which keeps generating $\bbot$.
Otherwise, $P$ is interpreted in $\bsigma \in \Sigma$, where a single rule at a time applies, whose premises interpreted from left to right produce states in the expected order.

\item
Theorem~\ref{theorem:forestReversibility} will formally show that the function in Definition~\ref{definition:Inverse of a term} generates the inverse of any $P\in\forestP$, according to the operational semantics.
\par
A simple example should help to follow how the inversion of an iteration works. Let
$P = \FF{from(i=-4 or}$ $\FF{0)to(i=1 or 0)\{j+=1\}}$ and $\bsigma \in \Sigma$ s.t. $\bsigma(\FF{i})=-4$ and $\bsigma(\FF{j}) = 2$.
Since \FF{-4<=1}, the interpretation of $P$ proceed forward, iterating $\FF{j+=1}$ until $\FF{i=1}$, making $|1 - (-4)| = 5$ iterations. The resulting state is therefore $\bsigmap = \bsigma[\FF{i}\!\mapsto\!1,\FF{j}\!\mapsto\!7]$.
Now, let us interpret the inverse $P^-$, which is \FF{from(i=1 or 0)to(i=-4 or 0)\{j+=1\}}, starting from $\bsigmap$. Since \FF{1>-4} we proceed backward, iterating \FF{j-=1} until $\FF{i=-4}$, making $|-4 - 1| = 5$ iterations yelding the starting state $\bsigma$.

\item
\label{remark: internalizing the menaing}
The rule \textsc{FromTo} is defined according to the idea of \emph{explaining the meaning} of \FF{from-to} by means of the terms that already belong to $\forestP$. Indeed, up to \FF{assert}, the unfolding of a loop relies on terms of \MSRL. A full internalization would require two further steps.
In analogy to \cite{MatosPR:RC20}, the first one would be to express $\FF{if(}e\FF{)\{}P\FF{\}else\{}Q\FF{\}}$ through \MSRL.
Secondly, it would be necessary to eliminate every \FF{assert}, possibly by leveraging partial evaluation techniques \cite{NormannGluck:PEPM24} or tools for static analysis \cite{ReholtGluckKruse:RC23,MalettoR:JLAMP24} developed for reversible computations. The consequence of such steps, would be to consider \MSRL as the core of $\forestP$, like \textsf{R-CORE} is for \textsf{R-WHILE} \cite{GluckYokoyama:TIS17}.
\qed
\end{enumerate}
\end{remark}
\section{\forest is terminating and reversible}
\label{section:Properties of Forest}
The goal of this section is twofold.
Firstly, it shows that, for every reasonable initial state, every $P \in \forest$ terminates either by failing or by yielding a state (Theorem~\ref{theorem:termination}). Then, it establishes that $P^-$ (Definition~\ref{definition:Inverse of a term}) truly represents the inverse of $P$, recovering the initial state from which $P$ may have been interpreted.
\par
The proofs of our goals will tacitly rely on the following fact, implied by the definition of the operational semantics:
\begin{fact}\label{fact:read-only-preservation}
Every non failing interpretation of $P\in\forest$ from a state $\bsigma$ produces a final state $\btau$, that preserves the read-only variables. Formally:
\begin{align*}
&\forall P\in\forest, \bsigma,\btau\in\Sigma.\ \bsigma\ P\ \btau \rightarrow (\forall x \notin (V\!\setminus\!\WDOM{P}). \ \btau(x) = \bsigma(x)) \enspace .\quad \qed
\end{align*}
\end{fact}

\begin{theorem}[Termination]\label{theorem:termination}
Let $P\in\forest$, and $\bsigma\in\Sigma$. The interpretation of $P$ starting from $\bsigma$, either terminates in some $\btau\in\Sigma$ or fails. Formally:
\[\forall P \in \forest, \forall \bsigma \in \Sigma.\ (\exists\btau \in \Sigma.\;\; \bsigma\ P\ \btau) \lor (\bsigma\ P\ \boldsymbol{\bot}) \enspace .\]
\end{theorem}
\noindent
The proof is by induction on the structure of $P$. A case worth commenting is when $P$ is $\FF{from(}i\FF{=}e_u\ \FF{or}\ e_{\operatorname{\textit{in}}}\FF{)to(}i\FF{=}e_v\  \FF{or}\  e_{\operatorname{\textit{out}}}\FF{)\{}Q\FF{\}}$.
\noindent
Let $\bsigma$ be an initial state. Let us assume that, initially, $\bsigma, e_u \FF{<=} e_v \Downarrow 1$.
The assertions at \Circled{0}, and \Circled{1} in Figure~\ref{fig: Computational flow of from-to forward} must be checked. If $\bsigma, e_u\FF{<=}i\ \FF{and}\ i\FF{<=}e_v \Downarrow 0$, namely $i$ is not in the interval delimited by $e_u, e_v$, or $\bsigma, e_{\operatorname{\textit{in}}} \Downarrow 0$ does not hold, then the iteration aborts, namely terminates, yielding $\bbot$.
Otherwise, the iteration starts, its termination depending on:
(a) $Q$ cannot write variable into $e_u$, $e_v$ nor $i$;
(b) by induction $Q$ terminates;
(c) every iteration at \Circled{3} in Figure~\ref{fig: Computational flow of from-to forward} increments $i$. Therefore, $i$ will at some point reach the upper bound determined by $e_v$, unless meanwhile, $e_{\operatorname{\textit{out}}}$ becomes true or $\FF{!}e_{\operatorname{\textit{in}}}$ false, allowing the loop to terminate even earlier. Furthermore the loop can also terminate earlier when $Q$ aborts or if the assertion at \Circled{3} fails.

Let us now assume $\bsigma, e_v \FF{<=} e_u \Downarrow 1$. Initially, $i$ must be into the interval determined by $e_v, e_u$. Termination works as above, but in point (c). Every iteration at \Circled{2} in Figure~\ref{fig: Computational flow of from-to backward} decrements $i$ which will reach the lower bound set by $e_v$ unless, meanwhile, $e_{\operatorname{\textit{out}}}$ becomes true or $\FF{!}e_{\operatorname{\textit{in}}}$ false. \qed

\begin{remark}[The interpretation may fail]
We stress that Theorem \ref{theorem:termination} only assures that the interpretation of a term $P$ terminates, but says nothing about its \emph{success} or \emph{failure}.
This is due to assertions. They are placed in strategic points to block the interpretation whenever a state that it produces does not assure that the interpreted term is reversible.
\end{remark}

\begin{theorem}[Reversibility]
\label{theorem:forestReversibility}
Let $P\in\forest$, and $\bsigma\in\Sigma$ an initial state. Then, $P$ generates a state $\btau$, namely $\bsigma\ P\ \btau$ if and only if $P^-$ generates $\bsigma$, starting from $\btau$. Formally:
$\forall P \in \forest, \forall \bsigma, \btau \in \Sigma.\ \bsigma\ P\ \btau \iff \btau\ P^-\ \bsigma $.
\end{theorem}

\begin{remark}
Theorem~\ref{theorem:forestReversibility} says that $P^-$ can recover the starting state $\bsigma$ exactly when the final state that $P$ generates from $\bsigma$ is not $\bbot$.
\qed
\end{remark}
\noindent
Clearly, one can simply say that \forest is reversible because it can be seen as a restriction of \JANUS. In any case, we briefly comment on how to prove Theorem~\ref{theorem:forestReversibility}. The proof splits into an `only if' and an `if' case, each proceeding by structural induction on the derivation with $\bsigma\ P\ \btau$, or $\btau\ P^-\ \bsigma$ as its conclusion, respectively.
\par
The main intuition follows.
Let $P \equiv \FF{from(}i\FF{=}e_u\ \FF{or}\ e_{\operatorname{\textit{in}}}\FF{)to(}i\FF{=}e_v\ \FF{or}\ e_{\operatorname{\textit{out}}}\FF{)\{}Q\FF{\}}$, and let $\bsigma \in \Sigma$. Let us focus on the case $\bsigma, e_u\FF{<=}e_v \Downarrow 1$, the other $\bsigma, e_u\FF{<=}e_v \Downarrow 0$ being analogous.
Rules \textsc{ForMain}, \textsc{IfTrue} and the repetition of \textsc{LoopRec} and \textsc{LoopBase} unfold the \FF{from-to} into a repetition of assertions, $P$ and the increment of $i$. Since, if $\bsigma \, P \, \btau$, we know how to prove:
\begin{align*}
	\bsigma \ &\FF{assert(}u\FF{<=}i \ \FF{and} \ i\FF{<=}v\FF{);assert(}i\FF{=}u \ \FF{or} \ e_{in}\FF{);}
	\\
	&[\FF{assert(!(}i\FF{=}v \ \FF{or} \ e_{out}\FF{));}P\FF{;}i\FF{+=1;assert(!(}i\FF{=}u \ \FF{or} \ e_{in}\FF{))}]^m \FF{;} \\
	&\FF{assert(}i \FF{=} v \ \FF{or} \ e_{out}\FF{)}\FF{;assert(}u \FF{<=} i \ \FF{and} \ i\FF{<=}v \FF{)} \ \btau \enspace ,
\end{align*}
for some $m \in \NN$, the resulting code can be inverted. Once inverted, it can be wrapped into $\FF{from(}i\FF{=}e_v\ \FF{or}\ e_{\operatorname{\textit{out}}}\FF{)to(}i\FF{=}e_u\ \FF{or}\ e_{\operatorname{\textit{in}}}\FF{)\{}Q\FF{\}}$ which is the inverse of $P$.

Concerning the selection, reversibility follows from the fact that its branches cannot alter the value of the guard.
\section{\forest is \PRF-complete because \MSRL is}
\label{section:forest is complete and sound with respect to MSRL}
Firstly, we briefly recall Matos' \MSRL from \cite{Matos:TCS03}.
Secondly, we provide Definition~\ref{definition:Embedding SRL}, a function translating \MSRL to \forest.
Theorem \ref{theorem:Soundness of forest w.r.t. MSRL} relies on it to prove that \forest is complete w.r.t. \MSRL.

\subsubsection*{Matos' \MSRL} stands for \emph{Simple Reversible Language} \cite{MATOS:TCS15}.
\MSRL programs read from, and write registers \FF{r}, \FF{r1}, \ldots, ranged over by $r$. Registers contain values in $\mathbb{Z}$.
Increment $\FF{INC} \ r$, and decrement $\FF{DEC}\ r$ of a register $r$ are the two basic programs of \MSRL, one inverse to the other, namely $\FF{INC} \ r$ is $(\FF{DEC} \ r)^{-}$, and vice versa.
If $P, Q$ are \MSRL programs with $P^{-}, Q^{-}$ their inverses, then \emph{series composition} $P\FF{;}Q$, and \emph{iteration} $\FF{for} \ r \ \FF{\{}P\FF{\}}$ are in \MSRL, their inverses being $Q^{-}\FF{;}P^{-}$, and $\FF{for} \ r \ \FF{\{}P^{-}\FF{\}}$.
If the value in $r$ is positive, an iteration is equivalent to $P\FF{;} \ldots \FF{;}P$ with as many $P$ as the value in $r$, register which cannot be written by $P$ If the value in $r$ is negative, an iteration is equivalent to $P^{-}\FF{;} \ldots \FF{;}P^{-}$ with as many $P^{-}$ as the \emph{inverse} of the value in $r$.
By definition, \MSRL has no \FF{if-then-else}. Using \MSRL terms \cite{MatosPR:RC20} encodes one in it.

\begin{remark}
Notions like ``domain/co-domain'', ``state'', etc.\@ defined for \forest obviously adapt to \MSRL. For example, $\bsigma \, P \, \btau$ stands for ``\MSRL program $P$ with a domain compatible with state $\bsigma$, produces a state $\btau$.''
Moreover, variables in \forest are registers in \MSRL.
\qed
\end{remark}

\subsubsection*{Translation.} The following notation eases the translation from \MSRL to \forest.

\begin{notation}[Leading variables]
If $P\in \forest$, the set $\lead{P}$ of \emph{leading} variables in $P$ contains all the variables \emph{leading} the unfolding of every \FF{from-to} in $P$. Let $\lead{Q}, \lead{Q'}$ be given for any $Q, Q'\in \forest$. Let $i\in V$, $e_u,e_v \in \EXPRZ$ and $e_{\operatorname{\textit{in}}}, e_{\operatorname{\textit{out}}} \in \EXPRB$. Then, $\lead{P}$ is empty for every $P\in\forest$ but the following two terms:
\begin{align}
\lead{Q\FF{;}Q'} & = \lead{Q} \cup \lead{Q'}
\\
\lead{
\FF{from(}i\FF{=}e_u\ \FF{or}\ e_{\operatorname{\textit{in}}}\FF{)}
\FF{to(}i\FF{=}e_v\ \FF{or}\ e_{\operatorname{\textit{out}}}\FF{)\{}Q\FF{\}}}
& = \{i\} \cup \lead{Q} \enspace .
\end{align}
\end{notation}

\noindent
Identifying the \emph{leading} variables avoids name clashes when an iteration of \MSRL is mapped to a \forest \FF{from-to}.

\begin{definition}[Translating \MSRL to \forest]
\label{definition:Embedding SRL}
Let $P, Q, Q' \in \MSRL$ and $r \in V$. The map $\STOF{\_}{}$ maps terms of \MSRL to terms in \forest as follows:
\begin{align*}
\STOF{\FF{INC} \ r}{} & = r \FF{+=}1 \qquad
\STOF{\FF{DEC} \ r}{} = r \FF{-=}1 \qquad
\STOF{Q\FF{;}Q'}{} = \STOF{Q}{}\FF{;}\STOF{Q'}{} \\[0.5\baselineskip]
\STOF{\FF{for} \ r \ \FF{\{}P\FF{\}}}{}
& = \FF{from(}i \FF{=} \FF{0}\ \FF{or}\ \FF{0)to(}i \FF{=} r\ \FF{or}\ \FF{0)\{} \STOF{P}{} \FF{\};} i \FF{-=} r
\enspace ,
\end{align*}
where
$i \in V$ is a \emph{fresh} variable not in $\{r\} \cup \DOM{\STOFI{P}} \cup \DOM{P}$, and
$\DOM{Q\FF{;}Q'}$ $\cap \lead{\STOFI{Q}} = \DOM{Q\FF{;}Q'}$ $\cap \lead{\STOFI{Q'}} =  \DOM{\FF{for} \ r \ \FF{\{}P\FF{\}}} \cap \lead{\STOFI{P}} = \emptyset$.
\end{definition}
\noindent
An example: $\STOFI{\FF{for r \{INC j\}}} =
\FF{from(i=0 or 0)to(i=r or 0)\{j+=1\};i-=r}$.

\subsubsection*{Completeness.} It is natural to require that $\STOFI{P}\in\forest$ \emph{simulates} any given $P\in\MSRL$, which means that $\forest$ is complete w.r.t. \MSRL.

\begin{notation}
Let $\bsigma, \bhsigma$ be states in $\Sigma$, let $P$ be a term in \MSRL or \forest.
\begin{enumerate}
\item
We say ``\emph{$\bsigma$ and $\bhsigma$ are equivalent w.r.t. P}'', written $\bsigma \simeq_P \bhsigma$, when  $\bsigma(r) = \bhsigma(r)$, for every $r \in \DOM{P}$, namely, if the two states have identical values on the variables of $P$.

\item
We say ``\emph{$\bsigma$ is $P$-clean}'', written $\clean{\bsigma}{P}$, when $\bsigma(x) = 0 $, for every $x\in \lead{P}$, namely when the state assigns $0$ to every leading variable of $P$.

\item
We say ``\emph{$\bhsigma$ simulates $\bsigma$ w.r.t. $P$}'', written $\bsigma \sqsubset_{P} \bhsigma$, if \textit{$\bsigma$ and $\bhsigma$ are equivalent w.r.t. $P$} and \emph{$\bsigma$ is $\STOFI{P}$-clean}.
\qed
\end{enumerate}
\end{notation}

\begin{theorem}[\forest is complete w.r.t. \MSRL]
\label{theorem:Soundness of forest w.r.t. MSRL}
Let $P\in\MSRL$, let $\bsigma,\btau\in\Sigma$ s.t. $\bsigma\,P\,\btau$. For any $\bhsigma$ s.t. $\bsigma \sqsubset_{P} \bhsigma$, we can derive $\bhsigma\, \STOFI{P}\, \bhtau$ for some $\bhtau$ s.t. $\btau \sqsubset_{P} \bhtau$.
\end{theorem}
\noindent
As for the proof, it is by induction on the structure of $P$, remarking that Definition~\ref{definition:Embedding SRL} assures $\DOM{\STOFI{P}} = \DOM{P} \cup \lead{\STOFI{P}}$ with $\DOM{P} \cap \lead{\STOFI{P}} = \emptyset$, for every \MSRL program $P$.
\par
We intuitively illustrate the case when $P$ is $\FF{for} \ r \ \FF{\{}Q\FF{\}}$, for some $r$ and $Q$. By definition, $\STOFI{\FF{for} \ r \ \FF{\{}Q\FF{\}}}$ yields
$\FF{from(}i\FF{=}0\ \FF{or}\ \FF{0)to(}i\FF{=}r \ \FF{or} \ \FF{0)\{}\STOFI{Q}\FF{\};}i\FF{-=}r$, where $i\notin \{r\} \cup \DOM{\STOFI{Q}}$ and $l(\STOFI{Q}) \cap \DOM{P} = \emptyset$.
Let $\bsigma, \btau \in \Sigma$ such that $\bsigma(r) = v \geq 0$ and $\bsigma \, P \, \btau$. The case with $v < 0$, is analogous.
In order to return $\btau$, $P$ executes $Q$ as many times as $v$, which we denote by $\bsigma \, [Q]^v \, \btau$.
Moreover, let $\bhsigma \in \Sigma$ s.t. $\bsigma \sqsubset_{P} \bhsigma$, hence $\bhsigma(i) = 0$, and let $\bhtau\in \Sigma$ s.t. $\bhsigma \, \STOFI{P} \, \bhtau$.
In order to produce $\bhtau$, it can be shown that the rules in Figure~\ref{fig:operationalSemantics} imply ``$\STOFI{P}$ executes `$\STOFI{Q}\FF{;}i\FF{+=1}$' $v$ times, followed by $i\FF{-=}r$'', which we shorten as $\bhsigma \, [\STOFI{Q}\FF{;}i\FF{+=1}]^v\FF{;}i\FF{-=}r \, \bhtau$.
\par
Now we are able to show that $\bhtau$ \textit{simulates $\btau$ w.r.t. $P$}, namely $\btau\sqsubset_P \bhtau$.
\par
\begin{enumerate}
\item
To ease the proof of the main statement, we can show what follows.
\par
By induction on $m$, for every $\bmu, \bnu, \bhmu$ s.t. $\bmu \sqsubset_Q \bhmu$, we can prove:
\[\bmu \ \, [Q]^m \ \, \bnu \implies \bhmu \ \, [\STOF{Q}{}\FF{;}i\FF{+=1}]^m \ \, \bhnu \]
where: (i) $\bnu \sqsubset_Q \bhnu$; (ii) $\bhnu(i) = \bhmu(i) + m$; and (iii) $\bhnu(r) = \bhmu(r)$, for some $\bhnu$.
\par
Specifically, knowing that, by definition, $i\not\in\DOM{\STOFI{Q}}$ we have that: (i) holds by applying the main statement, by induction, on $Q$;
(ii) holds because $i$ is incremented at every iteration;
(iii) holds because, by definition of \MSRL and the translating function,  $r\not\in\WDOM{Q}\cup\WDOM{\STOFI{Q}}$.

\item
Back to the main statement, we recall that, by assumption,  $\bsigma\ [Q]^v\ \btau $, and $\bsigma \sqsubset_P \bhsigma$ implies $\bsigma \sqsubset_Q \bhsigma$. So, the result in the previous point implies that:
$$\bhsigma\ [\STOF{Q}{}\FF{;}i\FF{+=1}]^v\ \bhtaup $$
where
(i) $\btau \sqsubset_Q \bhtaup$;
(ii) $\bhtaup(i) = \bhsigma(i) + m = 0 + m = m$;
and (iii) $\bhtaup(r) = \bhsigma(r) = m$,
for some $\bhtaup$.
\item
The conclusion $\btau \sqsubset_{P} \bhtaup$, follows by observing that it must be $\bhtau(i) = 0$, because  $\bhtau$ results from executing $i\FF{-=}r$ starting from $\bhtaup$.
\end{enumerate}

\begin{corollary}[\forest is \PRF-complete]
\label{corollary:forest is PRF-complete}
Let $f\in\PRF$. There exists $P\in\forest$ that computes $f$.
\end{corollary}
\noindent
As for the proof, \cite{MatosPR:RC20} implies that \MSRL is complete w.r.t. the class of \emph{Reversible Primitive Permutations} which in turn is \PRF-complete \cite{PaoliniPR:TCS20,MalettoR:RC22,MalettoR:JLAMP24}. By transitivity, Theorem~\ref{theorem:Soundness of forest w.r.t. MSRL} implies that \forest is \PRF-complete.

\begin{remark}
Since Corollary~\ref{corollary:forest is PRF-complete} shows that \forest is \PRF-complete, and Theorem~\ref{theorem:termination} implies that \forest is not Turing-complete, because it cannot develop infinite loops, we strongly believe that \forest is \emph{weakly \PRF-sound}, according to:
\begin{definition}[\forest is \emph{weakly} \PRF-sound]
For every $P\in \forest$, and $\bsigma, \btau\in\Sigma$, if $\bsigma\ P\ \btau$, namely $P$ does not fail, then a primitive recursive $\hat{P}\in\PRF$ exists that computes the same function as $P$.
\end{definition}
\noindent
We think that a proof of this would work in analogy to the one that \LOOP is \PRF-sound \cite{MeyerRitchie:ACM67}. We leave it to future work.
\qed
\end{remark}
\section{\forest is algorithmically more expressive than \MSRL}
\label{section:Algoritmic expressivity: SRL vs forest}
So far we know that \forest is \MSRL-complete, but we can also show:
\begin{theorem}
\label{theorem:forest more expressive than MSRL}
\forest is strictly algorithmically more expressive than \MSRL.
\end{theorem}
\noindent
We can prove it in three steps:
(i) we introduce a \forest term determining the \textit{sign} of an integer in time $O(1)$;
(ii) we introduce another \forest term computing the \textit{minimum} between two integers $m, n$ in $O(\MIN{|m|,|n|})$;
(iii) we argue about the non existence of a \MSRL program that can do same.

\subsubsection*{Computing the sign in \forest.}
We determine the sign of $n\in\ZZ$ in $O(1)$ steps, which is optimal, as follows.
\begin{definition}[Sign]
\label{definition:sign}
Let \FF{sign} be the name for:
\[
\FF{from} \, \FF{(}i\FF{=0} \ \FF{or} \ \FF{0)} \, \FF{to} \,  \FF{(}i\FF{=}x \ \FF{or} \ \FF{!(}s\FF{=0))} \, \FF{\{}s\FF{+=1\}}
\]
where $x$ is the variable we want the sign of, and $i, s$ the variables eventually containing the sign.
\end{definition}
\noindent
Starting from $\bsigma\in\Sigma$ s.t. $\bsigma(i) = 0$ and $\bsigma(s) = 0$,
we can see that $\bsigma\ \FF{sign}\ \bsigma[i \mapsto v, s \mapsto v]$, where:
\[
v =
\begin{cases}
 1  &\text{if \xspace$\bsigma(x) > 0$}\\
 0  &\text{if \xspace$\bsigma(x) = 0$} \\
-1  &\text{if \xspace$\bsigma(x) < 0$}
\end{cases}
\enspace .
\]
\noindent
The key point of \FF{sign} is to fully exploit the interpretation given by the rule \textsc{FromTo} which interprets $\FF{\{}s\FF{+=1\}}$, skip it, or interprets its inverse, according to the value in $x$, which can be greater, equal, or greater than $0$, respectively.

\begin{fact}
Both time and space complexities of \FF{sign} are in $O(1)$.
\qed
\end{fact}
\noindent
The behavior of \FF{sign} can be traced back to how \cite{MatosPR:RC20} encodes the sign function inside \SRL. However, \MSRL iterations cannot be preemptively interrupted once the sign has been discovered, while \forest ones can. Due to this, the sign algorithm in \cite{MatosPR:RC20} has quadratic time and exponential space complexity.
Thus, \FF{sign} in Definition \ref{definition:sign} is a first example of how ``\textit{escaping} from loops'' extends \forest algorithmic expressiveness as compared to \MSRL, possibly reducing drastically time complexity.

\begin{lstlisting}[style=forest, caption=Minimum between two negative values, label=listing:minimum between two negatives]
// (*$m, n \in \ZZ^-$*), x=(*$m$*), y=(*$n$*), i=0, min=0, found=0
min += y;
from ((i=0) or 0) to ((i=-x) or (found=1)) {
  if (i=-y) { // |y| < |x| -> x < y
    min -= y;
    min += x;
    found += 1
  } else {skip}
}
\end{lstlisting}

\subsubsection*{Computing the minimum in \forest.}
We detail out the main ideas to write a \forest term \FF{minGen} that computes the minimum between any $m, n\in\ZZ$ in time $O(\min{|m|,|n|})$.
\par
Let \FF{x}, \FF{y} be two variables holding $m$ and $n$, respectively.
Let \FF{min}, \FF{i}, \FF{found} be output and auxiliary variables, initially set to $0$.
The term \FF{minGen} must start with two instances of \FF{sign} to determine the sign of \FF{x}, and \FF{y}.
Once the sign is known, \FF{minGen} can distinguish among four scenarios:
\begin{enumerate}
\item Both \FF{x}, and \FF{y} are positive. In this case \FF{minGen} must behave as \FF{minPos} in Listing \ref{listing:Minimum minimumF in forest}, which works as follows. It assumes that $m$ in \FF{x} is the minimum, setting \FF{min} to it. Then, \FF{i} counts from $0$ to \FF{x}. If \FF{i} reaches \FF{y} before getting to \FF{x}, it means that \FF{y} is the least value. So, \FF{minGen} sets \FF{min} to \FF{y}, stopping the iteration by setting \FF{found} to $1$. Otherwise, \FF{min} stays at \FF{x}.
\item Only one between \FF{x}, and \FF{y} is positive. Then, \FF{minGen} must set $\FF{min}$ to the variable containing the negative number, and this is trivial.
\item  Both \FF{x}, and \FF{y} are negative. Then, \FF{minGen} must behave as \FF{minNeg} in Listing \ref{listing:minimum between two negatives}, analogous to \FF{minPos}, but working on the absolute values of \FF{x}, and \FF{y}.
\end{enumerate}
\noindent
Since, by construction, \FF{sign} costs $O(1)$, and every case is in $O(\MIN{|m|,|n|})$, the whole \FF{minGen} costs $O(\MIN{|m|,|n|})$. Therefore, we can state:

\begin{proposition}
\label{proposition:optimal minimum in forest}
There exists a term in \forest computing the minimum between $m,n\in\ZZ$ in time $O(\MIN{|m|,|n|})$.
\end{proposition}

\subsubsection*{\MSRL has no optimal minimum algorithm.}
We devise this proof following Matos. He proves an analogous of \emph{ultimate obstinacy property} for \LOOP languages \cite{MATOS:TCS15}. Intuitively, we recall that for a computational model being \emph{ultimately obstinate} means that some functions cannot be computed by its programs in optimal time because such functions are \emph{non trivial}, namely they \emph{cannot} have form $f(x_1,\ldots,x_n) = x_i + c$ \cite{Colson:TCS98}.

Clearly, the \textit{minimum} between $m, n$ is non trivial, thus every program in \MSRL implementing it
must be subject to two structural constraints.
They must contain two registers, say \FF{rm}, \FF{rn} storing $m, n$ respectively. Moreover, they must have form $P_1\, \FF{;}\, \FF{for}\, r\, \{P_2\}\, \FF{;}\, P_3$ where $r$ has to be one between \FF{rm} and \FF{rn}.

Let us assume that $r$ is \FF{rm}, the alternative choice being equivalent.
For our purposes \emph{the structure} of $P_2$ and $P_3$ is irrelevant, while the structure of $P_1$ is.
$P_1$ can be empty, an explicit sequence of increments/decrements with fixed length, or even contain iterations, none of them led by \FF{rm} or \FF{rn}.
Therefore, once we start interpreting $\FF{for rm \{}P_2\FF{\}}$, due to $P_1$ structure, \FF{rm} contains $m + c$, for some constant $c\in\ZZ$, implying that the time complexity of $\FF{for rm \{}P_2\FF{\}}$ is at least $O(m)$. Clearly, if $r$ were \FF{rn}, the time complexity would be $O(n)$. This implies:

\begin{proposition}
\label{proposition:no optimal minimum in MSRL}
No $\MSRL$ program exists that computes the minumum between $m, n \in \ZZ$ in time $O(\MIN{|m|,|n|})$.
\end{proposition}
\noindent
So, both Propositions \ref{proposition:optimal minimum in forest}, and  \ref{proposition:no optimal minimum in MSRL} imply Theorem \ref{theorem:forest more expressive than MSRL}.
\section{Conclusions and future work}
\label{section:Conclusions, future work}
We addressed the problem of algorithmic expressiveness in a reversible setting.
Our artifact is \forest, a reversible computational model where iterations always terminate, but can be interrupted preemptively. \forest is a meeting point between the two established and reversible computational models \MSRL and \JANUS.
\forest is \MSRL and \PRF-complete, but strictly more algorithmically expressive. It allows us to write an algorithm that computes the minimum between two integers $m$ and $n$ in $O(\MIN{m,n})$, which cannot be encoded in \SRL.
The reason is that we can stop \forest iterations as soon as the expected result is available, while \MSRL iterations always unfold to the end.

\paragraph{Future work.}
Even though \forest terms always terminate, this can be due to failing assertions. Clearly, assertions seems to be necessary for loops to have general exit condition. However, failing computations cannot be reversed. We aim to remove assertions to produce a computational model with same algorithmic expressiveness as \forest, whose programs always terminate successfully. A possible strategy to follow is at least \cite{ReholtGluckKruse:RC23}, where authors rely on \textsf{SMT} solvers to statically remove unnecessary assertions.

A further step we plan is to compare our work with the one of \cite{ValarcherAndaryPatrou:FI11}, which, in fact, inspired us.
The authors of \cite{ValarcherAndaryPatrou:FI11} introduce \loopExit. It is an \textit{unstructured} conservative extension of \LOOP models. The extension consists of introducing a statement \LL{exit} which can break loops by jumping to end of a program.
The main result in \cite{ValarcherAndaryPatrou:FI11} is that \loopExit is as algorithmically expressive as \textsf{APRA}, a \PRF-sound and complete computational model obtained by restricting Gurevich \textit{Abstract State Machines}  \cite{DBLP:conf/asm/Gurevich93}.
In fact, in relation to the algorithmic expressiveness, we think that the most valuable goal would be to introduce a reversible
version of \textsf{APRA} subsuming \forest.

\printbibliography
\end{document}